\documentstyle[preprint,aps,epsf]{revtex}
\begin{document}

\author{A. B\'{e}rard$^{a}$, H. Mohrbach$^{a}$ and P. Gosselin$^{b}$. \\
\textit{a)L.P.L.I. Institut\ de\ Physique,}\\
\textit{\ 1 blvd.F.Arago, 57070, Metz, France.}\\
\textit{b)Universit\'{e} Grenoble I , Institut Fourier, }\\
\textit{UMR 5582 CNRS, UFR de Math\'{e}matique, BP\ 74, }\\
\textit{38402 Saint Martin d'H\`{e}res, Cedex, France.}}
\title{Lorentz-Covariant Hamiltonian Formalism. }
\maketitle

\begin{abstract}
The dynamic of a classical system can be expressed by means of Poisson
brackets. In this paper we generalize the relation between the usual non
covariant Hamiltonian and the Poisson brackets to a covariant Hamiltonian
and new brackets in the frame of the Minkowski space. These brackets can be
related to those used by Feynman in his derivation of Maxwell's equations. 
The case of curved space is also considered with the introduction of Christoffel
symbols, covariant derivatives, and curvature tensors.
\end{abstract}

\section{Introduction}

A remarkable formulation of classical dynamics is provided by 
Hamiltonian mechanics. This is an old subject. However, new
discoveries are still been made; we quote two examples among several: 
the Arnold duality transformations, which generalize the canonical
transformations \cite{ARNOLD,STUMP}, and the extensions of the Poisson
brackets to differential forms and multi-vector fields by A.Cabras and
M.Vinogradov \cite{CABRAS}. In this context the transition from classical
to relativistic mechanics raises the question of Hamiltonian
covariance, the physical significance of which is discussed by Goldstein 
\cite{GOLDSTEIN}.
In the first part of this paper we briefly recall the Poisson brackets
approach and the covariant Hamiltonian formalism. Then we introduce new 
brackets to study the dynamics associated to this covariant
Hamiltonian, which define an algebraic structure between position and
velocity, and does not have an explicit formulation. We examine
the close link between these brackets and those used by Feynman for his
derivation of the Maxwell equations \cite{DYSON,TANIMURA,NOUS1,NOUS2}. 
A very interesting way to arrive at the same sort of result was found by 
Souriau in the frame of his symplectic classical mechanics \cite
{SOURIAU}. In the final part of this work we consider the dynamics in 
curved space, using Christoffel symbols, covariant derivatives, and curvature
tensors expressed in terms of these brackets.

\section{Brief review of Analytic Mechanics}

\subsection{Poisson brackets}

The dynamics of a classical particle in a 3-dimensional flat space with
vector position $q^{i}$ and vector momentum $p_{i}$ ($i=1,2,3)$ is defined
by the Hamilton equations:

\begin{equation}
\left\{ 
\begin{array}{c}
\stackrel{.}{q}^{i}=\frac{dq^{i}}{dt}=\frac{\partial H}{\partial p_{i}} \\ 
\\ 
\stackrel{.}{p}_{i}=\frac{dp_{i}}{dt}=-\frac{\partial H}{\partial q^{i}}
\end{array}
\right.
\end{equation}
where the Hamiltonian $H(q^{i},p_{i})$ is a form on the phase space ( the
cotangent fiber space). They can be also expressed in a symmetric manner by
means of Poisson brackets:

\begin{equation}
\left\{ 
\begin{array}{c}
\stackrel{.}{q}^{i}=\left\{ q^{i},H\right\} \\ 
\\ 
\stackrel{.}{p}_{i}=\left\{ p_{i},H\right\}
\end{array}
\right.
\end{equation}
These brackets are naturally defined as skew symmetric bilinear maps on the
space of functions on the phase space in the following form:

\begin{equation}
\left\{ f,g\right\} =\frac{\partial f}{\partial q^{i}}\frac{\partial g}{%
\partial p_{i}}-\frac{\partial g}{\partial q^{i}}\frac{\partial f}{\partial
p_{i}}
\end{equation}

\subsection{Covariant Hamiltonian}

Except in the electromagnetic situation, the Hamiltonian is not the total
energy when it is time-dependent, and its generalization to 
relativistic problems with the $M_{4}$ Minkowski space is not trivial
because it is not Lorentz covariant.

In the electromagnetic case the answer to this question is given by the
introduction of the following covariant expression\cite{GOLDSTEIN} :

\begin{equation}
H=u^{\mu }p_{\mu }-L=u^{\mu }(mu_{\mu }+\frac{q}{c}A_{\mu })\text{ }
\end{equation}
where $L$ is the usual invariant electromagnetic Lagrangian : 
\begin{equation}
L=\frac{1}{2}m\text{ }u^{\mu }u_{\mu }+\frac{q}{c}\text{ }u^{\mu }A_{\mu }
\end{equation}
and $u^{\mu }$ the quadri-velocity defined by means of the proper time $
t_{p} $, here used as an invariant parameter:

\begin{equation}
u^{\mu }=\frac{dx^{\mu }}{dt_{p}}
\end{equation}
Finally we have the covariant Hamiltonian

\begin{equation}
H=\frac{1}{2}m\text{ }u^{\mu }u_{\mu }
\end{equation}
with the corresponding eight Hamilton equations:

\begin{equation}
\left\{ 
\begin{array}{c}
\frac{\partial H}{\partial p_{\mu }}=\frac{dx^{\mu }}{dt_{p}}=u^{\mu } \\ 
\\ 
\frac{\partial H}{\partial x^{\mu }}=-\frac{dp_{\mu }}{dt_{p}}
\end{array}
\right.
\end{equation}
It is interesting to recall that this structure is only possible in the
situation where the potential can be put in a covariant manner as in the
electromagnetism theory.

\section{Lorentz covariant Hamiltonian and brackets formalism}

Now we want to generalize the relation between the usual non covariant
relativistic Hamiltonian and the Poisson brackets to a covariant Hamiltonian 
$H$ and new formal brackets introduced in the frame of the
Minkowski space. It is important to remark that, in a different manner,
P.Bracken also studied the relation between this Feynman problem and
the Poisson brackets\cite{BRACKEN}.

In this context a ''dynamic evolution law'' is given by means of a one real
parameter group of diffeomorphic transformations :

\[
g\text{ }(\text{ }IR\text{ }\times \text{ }M_{4}\text{ })\longrightarrow
M_{4}:\text{ }g(\tau ,x)=g^{\tau }x=x(\tau ) 
\]
The ''velocity vector'' associated to the particle is naturally introduced
by:

\begin{equation}
\stackrel{.}{x}^{\mu }=\frac{d}{d\tau }g^{\tau }x^{\mu }
\end{equation}
where the ''time'' $\tau $ is not identified with the proper time as we
shall see later. The derivative with respect to this parameter of an arbitrary 
function defined on the tangent bundle space can be written, by means of the
covariant Hamiltonian, as:

\begin{equation}
\frac{df(x,\stackrel{.}{x},\tau )}{d\tau }=\left[ H,\text{ }f(x,\stackrel{.}{%
x},\tau )\right] +\frac{\partial f(x,\stackrel{.}{x,\tau })}{\partial \tau }
\label{hamilton}
\end{equation}
where for $H$ we take the following definition:

\begin{equation}
H=\frac{1}{2}\text{ }m\text{ }\frac{dx^{\mu }}{d\tau }\text{ }\frac{dx_{\mu }%
}{d\tau }=\frac{1}{2}\text{ }m\stackrel{.}{x}^{\mu }\stackrel{.}{x}_{\mu }
\end{equation}
\smallskip 
Equation (\ref{hamilton}) giving the dynamic of the system, is
the definition of our new brackets structure, and is the fundamental
equation of the paper.

We require for these new brackets the usual first Leibnitz law:

\begin{equation}
\left[ A,BC\right] =\left[ A,B\right] C+\left[ A,C\right] B
\end{equation}
and the skew symmetry:

\begin{equation}
\left[ A,B\right] =-\left[ B,A\right]
\end{equation}
where the quantities $A,$ $B$ and $C$ depend of $x^{\mu }$ and $\stackrel{.}{%
x}^{\mu }$.

In the case of the vector position $x^{\mu }(\tau )$ we have from (\ref
{hamilton}):

\begin{equation}
\stackrel{.}{x}^{\mu }=\left[ H,x^{\mu }\right] =m\text{ }\left[ \stackrel{.%
}{x}^{\nu },x^{\mu }\right] \stackrel{.}{x}_{\nu }
\end{equation}
and we easily deduce that:

\begin{equation}
m\text{ }\left[ \stackrel{.}{x}^{\nu },x^{\mu }\right] =g^{\mu \nu }
\label{métrique}
\end{equation}
where $g^{\mu \nu }$ is the metric tensor of the Minkowski space.

As in the Feynman approach the time parameter is not the proper time. To see
this we borrow Tanimura's argument \cite{TANIMURA}. Consider the relation 
\begin{equation}
g^{\mu \nu }\frac{dx^{\mu }}{dt_{p}}\frac{dx^{\nu }}{dt_{p}}=1
\end{equation}
\smallskip which implies 
\begin{equation}
\left[ \stackrel{.}{x}^{\lambda },g^{\mu \nu }\frac{dx^{\mu }}{dt_{p}}\frac{%
dx^{\nu }}{dt_{p}}\right] =0.
\end{equation}
and is in contradiction with: 
\begin{equation}
\left[ \stackrel{.}{x}^{\lambda },g^{\mu \nu }\frac{dx^{\mu }}{d\tau }\frac{%
dx^{\nu }}{d\tau }\right] =-\frac{2}{m}\stackrel{.}{x}^{\lambda }.
\end{equation}
But differently from Feynman, the fact that $g^{\mu \nu }$ is the metric is
a consequence of the formalism and is not imposed by hand. In addition,
contrary to Feynman, we do not need to impose the Leibnitz condition:

\begin{equation}
\frac{d}{d\tau }\left[ A,B\right] =\left[ \frac{dA}{d\tau },B\right] +\left[
A,\frac{dB}{d\tau }\right]  \label{leib}
\end{equation}
($A$ and $B$ being position- and velocity-dependent functions) because the
time derivative is given by the fundamental equation (\ref{hamilton}).

We impose the usual locality property:

\begin{equation}
\left[ x^{\mu },x^{\nu }\right] =0
\end{equation}
which directly gives for an expandable function of the position or the
velocity the following useful relations:

\begin{equation}
\left\{ 
\begin{array}{c}
\left[ x^{\mu },f(\stackrel{.}{x})\right] =-\frac{1}{m}\frac{\partial f(%
\stackrel{.}{x})}{\partial \stackrel{.}{x}_{\mu }} \\ 
\\ 
\left[ \stackrel{.}{x}^{\mu },f(x)\right] =\frac{1}{m}\frac{\partial f(x)}{%
\partial x_{\mu }}
\end{array}
\right.
\end{equation}
which reduce in the particular cases of the position and velocity to: 
\begin{equation}
\left\{ 
\begin{array}{c}
\left[ x^{\mu },\stackrel{.}{x}^{\nu }\right] =-\frac{1}{m}g^{\mu \rho }%
\frac{\partial \stackrel{.}{x}^{\nu }}{\partial \stackrel{.}{x}^{\rho }}=-%
\frac{1}{m}\frac{\partial \stackrel{.}{x}^{\nu }}{\partial \stackrel{.}{x}%
_{\mu }}=-\frac{g^{\mu \nu }}{m} \\ 
\\ 
\left[ \stackrel{.}{x}^{\mu },x^{\nu }\right] =\frac{1}{m}g^{\mu \rho }\frac{%
\partial x^{\nu }}{\partial x^{\rho }}=\frac{1}{m}\frac{\partial x^{\nu }}{%
\partial x_{\mu }}=\frac{g^{\mu \nu }}{m}
\end{array}
\right.  \label{GMUNU}
\end{equation}
To compute the bracket between two components of the velocity we require in
addition the Jacobi identity:

\begin{equation}
\left[ \left[ \stackrel{.}{x}^{\mu },\stackrel{.}{x}^{\nu }\right] ,x^{\rho
}\right] +\left[ \left[ x^{\rho },\stackrel{.}{x}^{\mu }\right] ,\stackrel{.%
}{x}^{\nu }\right] +\left[ \left[ \stackrel{.}{x}^{\nu },x^{\rho }\right] 
\stackrel{.}{x}^{\mu }\right] =0  \label{jacobi1}
\end{equation}
which by using (\ref{métrique}) gives:

\begin{equation}
\left[ \stackrel{.}{x}^{\mu },\stackrel{.}{x}^{\nu }\right] =-\frac{N^{\mu
\nu }(x)}{m}
\end{equation}
where $N^{\mu \nu }(x)$ is a skew symmetric tensor.

The second derivative of the position vector is:

\begin{equation}
\stackrel{..}{x}^{\mu }=\frac{d\stackrel{.}{x}^{\mu }}{d\tau }=\left[ H,%
\stackrel{.}{x}^{\mu }\right] =N^{\mu \nu }\stackrel{.}{x}_{\nu }
\label{motion2}
\end{equation}
and we write:

\begin{equation}
F^{\mu \nu }=\frac{m}{q}N^{\mu \nu }
\end{equation}
in order to recover the Lorentz equation of motion.

\textit{remark 1.} We can easily calculate:  
\begin{equation}
\left[ H,H\right] =\frac{1}{4}m^{2}\left[ \stackrel{.}{x}_{\mu }\stackrel{.}{%
x}^{\mu },\stackrel{.}{x}_{\nu }\stackrel{.}{x}^{\nu }\right] =-\text{ }q%
\stackrel{.}{x}_{\mu }\stackrel{.}{x}_{\nu }F^{\mu \nu }=0
\end{equation}
and then deduced:  
\begin{equation}
\frac{dH}{d\tau }=\frac{\partial H}{\partial \tau }
\end{equation}
which is the expected result.

In the same manner, we get for the $4$-orbital momentum: 
\begin{eqnarray}
\frac{dL^{\mu \nu }}{d\tau } &=&m\frac{d}{d\tau }\left( x^{\mu }\stackrel{.}{%
x}^{\nu }-\stackrel{.}{x}^{\mu }x^{\nu }\right) =m\left( x^{\mu }\stackrel{..%
}{x}^{\nu }-\stackrel{..}{x}^{\mu }x^{\nu }\right)  \nonumber \\
&=&q(x^{\mu }F^{\nu \rho }\stackrel{.}{x}_{\rho }-x^{\nu }F^{\mu \rho }%
\stackrel{.}{x}_{\rho })=\left[ H,L^{\mu \nu }\right]
\end{eqnarray}
as expected.

\section{Maxwell equations}

Our formal construction will give the Maxwell equations because it leads to
the fundamental result (\ref{métrique}) which is the starting point of
Feynman's proof of the first group of Maxwell equations. The difference is
that our main property is equation (\ref{hamilton}) and not the Leibnitz
rule (\ref{leib}). So our derivation will be obtained differently and will
give in addition the two groups of Maxwell equations.

$\bullet $ To be general, we choose like in \cite{NOUS2}, the following
definition for the gauge curvature: 
\begin{equation}
\left[ \stackrel{.}{x^{\mu }},\stackrel{.}{x^{\nu }}\right] =-\text{ }\frac{1%
}{m^{2}}(qF^{\mu \nu }+g^{*}F^{\mu \nu })  \label{qq}
\end{equation}
where g will be interpreted as the magnetic charge of the Dirac monopole,
the *-operation being the Hodge duality.

$\bullet $ A simple derivative gives:

\begin{equation}
\frac{d(qF^{\mu \nu }(x)+g^{*}F^{\mu \nu }(x))}{d\tau }=q\partial ^{\rho
}F^{\mu \nu }(x)\stackrel{.}{x}_{\rho }+g\partial ^{^{\rho }*}F^{\mu \nu }(x)%
\stackrel{.}{x}_{\rho }  \label{dfdt1}
\end{equation}
and by means of the fundamental relation (\ref{hamilton}) we
obtain: 
\begin{eqnarray}
\frac{d(qF^{\mu \nu }(x)+g^{*}F^{\mu \nu }(x))}{d\tau } &=&\left[ H,qF^{\mu
\nu }(x)+g^{*}F^{\mu \nu }(x)\right]  \nonumber \\
&=&-\frac{m^{3}}{q}\text{ }\left[ \stackrel{.}{x}^{\rho },\left[ \stackrel{.%
}{x}^{\mu },\stackrel{.}{x}^{\nu }\right] \right] \stackrel{.}{x}_{\rho }
\end{eqnarray}
Now using the Jacobi identity we rewrite this expression as:

\begin{eqnarray}
\frac{d(qF^{\mu \nu }(x)+g^{*}F^{\mu \nu }(x))}{d\tau } &=&\frac{m^{3}}{q}%
\left( \left[ \stackrel{.}{x}^{\mu },\left[ \stackrel{.}{x}^{\nu },\stackrel{%
.}{x}^{\rho }\right] \right] \stackrel{.}{x}_{\rho }+\text{ }\left[ 
\stackrel{.}{x}^{\nu },\left[ \stackrel{.}{x}^{\rho },\stackrel{.}{x}^{\mu
}\right] \right] \stackrel{.}{x}_{\rho }\right) \stackrel{.}{x}_{\rho } 
\nonumber \\
&=&-q(\partial ^{\mu }F^{\nu \rho }+\text{ }\partial ^{\nu }F^{\rho \mu }x)%
\stackrel{.}{x}_{\rho }-g(\partial ^{\mu }{}^{*}F^{\nu \rho }+\text{ }%
\partial ^{\nu }{}^{*}F^{\rho \mu }x)\stackrel{.}{x}_{\rho }  \nonumber \\
&&  \label{dfdt3}
\end{eqnarray}
By comparing equations (\ref{dfdt1}) and (\ref{dfdt3}) we deduce the
following field equation:

\begin{equation}
q(\partial ^{\mu }F^{\nu \rho }+\partial ^{\nu }F^{\rho \mu }+\partial
^{\rho }F^{\mu \nu })+g(\partial ^{\mu }{}^{*}F^{\nu \rho }+\partial ^{\nu
}{}^{*}F^{\rho \mu }+\partial ^{\rho }{}^{*}F^{\mu \nu })=0
\end{equation}
that is: 
\begin{equation}
\left\{ 
\begin{array}{c}
\partial ^{\mu }F^{\nu \rho }+\partial ^{\nu }F^{\rho \mu }+\partial ^{\rho
}F^{\mu \nu }=gN^{\mu \nu \rho } \\ 
\\ 
\partial ^{\mu }{}^{*}F^{\nu \rho }+\partial ^{\nu }{}^{*}F^{\rho \mu
}+\partial ^{\rho }{}^{*}F^{\mu \nu }=-qN^{\mu \nu \rho }
\end{array}
\right.
\end{equation}
where $N^{\mu \nu \rho }$ is a tensor to be interpreted.

Using the differential forms language defined on the Minkowski space $%
(M_{4}) $ we write the preceding equations in a compact form: 
\begin{equation}
\left\{ 
\begin{array}{c}
dF=gN \\ 
\\ 
d^{*}F=-qN
\end{array}
\right.
\end{equation}
where $F$ and $^{*}F\in $ $\wedge ^{2}(M_{4})$ and $N$ $\in \wedge
^{3}(M_{4})$ .

If we put: 
\begin{equation}
\left\{ 
\begin{array}{c}
gN=-^{*}k \\ 
\\ 
qN=^{*}j
\end{array}
\right.
\end{equation}
where $j$ and $k$ $\in \wedge ^{1}(M_{4})$, we deduce: 
\begin{equation}
\left\{ 
\begin{array}{c}
\delta F=j \\ 
\\ 
dF=-^{*}k
\end{array}
\right.
\end{equation}
$\delta $ is the usual codifferential

\[
\delta :\wedge ^{k}(M_{4})\rightarrow \wedge ^{k-1}(M_{4}) 
\]
defined here as:

\[
\delta =(-)^{k(4-k+1)+1}(^{*}d^{*}) 
\]
Interpreting the $1$-forms $j$ and $k$ as the electric and magnetic four
dimensional current densities, we obtained the two groups of Maxwell
equations in the presence of a magnetic monopole. The situation without
monopole is evidently obtained by putting the $1$-form $k$ equal to zero.

We easily see by means of the Poincar\'{e} theorem that:

\begin{equation}
\delta ^{2}F=\delta j=0
\end{equation}
which is nothing else that the current density continuity equation:

\begin{equation}
\partial _{\mu }\text{ }j^{\mu }=m\left[ \stackrel{.}{x}_{\mu },\text{ }%
j^{\mu }\right] =0,
\end{equation}
From the skew property of the brackets, we can choose:

\begin{equation}
j^{\mu }=\rho \stackrel{.}{x}^{\mu },
\end{equation}
$\rho $ is the charge density whose dynamic evolution is
given by:

\begin{equation}
\frac{d\rho }{d\tau }=\left[ H,\rho \right] =m\left[ \stackrel{.}{x}^{\mu
},\rho \right] \stackrel{.}{x}_{\mu }=\left( \partial ^{\mu }\rho \right) 
\stackrel{.}{x}_{\mu }=\partial ^{\mu }j_{\mu }=0
\end{equation}
We see that $H$ automatically takes into account the gauge curvature. It
plays the role of a Hamiltonian not with the usual Poisson brackets, but
with new four-dimensional brackets which can be related to for example, 
those used by Feynman in his derivation of Maxwell equations as published 
by Dyson \cite{DYSON}.

\section{Application to a curved space}

In this section we extend the previous analysis to the case of a general
space time metric $g_{\mu \nu }(x)$.

In this case we define the covariant Hamiltonian from the usual fundamental
quadratic form $ds^{2}$ in the following manner: 
\begin{eqnarray*}
H &=&\frac{1}{2}m\left( \frac{ds}{d\tau }\right) ^{2}=
\frac{1}{2}mg_{\mu \nu }(x)\stackrel{.}{x}^{\mu }\stackrel{.}{x}^{\nu }
\end{eqnarray*}
In the same manner asin section $3$, we can prove the relation between the
metric tensor and the brackets structure: 
\[
m\text{ }\left[ \stackrel{.}{x}^{\nu },x^{\mu }\right] =g^{\mu \nu }(x) 
\]
The law of motion is:

\begin{eqnarray}
\stackrel{..}{x}^{\mu } &=&\left[ H,\stackrel{.}{x}^{\mu }\right] =\frac{1}{2%
}m\left[ g_{\nu \rho },\stackrel{.}{x}^{\mu }\right] \stackrel{.}{x}^{\nu }%
\stackrel{.}{x}^{\rho }+m\left[ \stackrel{.}{x}^{\nu },\stackrel{.}{x}^{\mu
}\right] \stackrel{.}{x}_{\nu }  \nonumber \\
&=&-\frac{1}{2}\text{ }\partial ^{\mu }g_{\nu \rho }\stackrel{.}{x}^{\nu }%
\stackrel{.}{x}^{\rho }-N^{\nu \mu }\stackrel{.}{x}_{\nu }  \label{motion3}
\end{eqnarray}
where we define $N^{\mu \nu }(x,\stackrel{.}{x})$ as:

\begin{equation}
\left[ \stackrel{.}{x}^{\mu },\stackrel{.}{x}^{\nu }\right] =-\frac{N^{\mu
\nu }(x,\stackrel{.}{x})}{m}
\end{equation}
Note that this tensor is now velocity-dependent, in contrast to the Minkowski
case.

By means of equation (\ref{jacobi1}) and (\ref{motion3}), we deduce the
relation:

\begin{equation}
\frac{\partial N^{\mu \nu }}{\partial \stackrel{.}{x}_{\rho }}=\partial
^{\nu }g^{\rho \mu }-\partial ^{\mu }g^{\rho \nu }
\end{equation}
then:

\begin{equation}
N^{\mu \nu }(x,\stackrel{.}{x})=-\text{ }(\partial ^{\mu }g^{\rho \nu
}-\partial ^{\nu }g^{\rho \mu })\stackrel{.}{x}_{\rho }+n^{\mu \nu }(x)
\end{equation}
where the tensor $n^{\mu \nu }(x)$ is only position dependent. If we
introduce this equation in (\ref{motion3}), we find:

\begin{eqnarray}
\stackrel{..}{x}^{\mu } &=&-\frac{1}{2}\text{ }\partial ^{\mu }g_{\nu \rho }%
\stackrel{.}{x}^{\nu }\stackrel{.}{x}^{\rho }-(\partial ^{\mu }g^{\rho \nu
}-\partial ^{\nu }g^{\rho \mu })\stackrel{.}{x}_{\nu }\stackrel{.}{x}_{\rho
}+n^{\mu \nu }(x)x_{\nu }  \nonumber \\
&=&\frac{1}{2}\text{ }\partial ^{\mu }g^{\nu \rho }\stackrel{.}{x}_{\nu }%
\stackrel{.}{x}_{\rho }-(\partial ^{\mu }g^{\rho \nu }-\frac{1}{2}\partial
^{\nu }g^{\rho \mu }-\frac{1}{2}\partial ^{\rho }g^{\nu \mu })\stackrel{.}{x}%
_{\nu }\stackrel{.}{x}_{\rho }+n^{\mu \nu }(x)\stackrel{.}{x}_{\nu } 
\nonumber \\
&=&-\Gamma ^{\nu \rho \mu }\stackrel{.}{x}_{\nu }\stackrel{.}{x}_{\rho
}+n^{\mu \nu }(x)\stackrel{.}{x}_{\nu }
\end{eqnarray}
where we have defined the Christoffel symbols by:

\begin{eqnarray}
\Gamma ^{\nu \rho \mu } &=&\frac{1}{2}\text{ }\left( \left[ \stackrel{.}{x}%
^{\rho },\left[ \stackrel{.}{x}^{\nu },x^{\mu }\right] \right] -\left[ 
\stackrel{.}{x}^{\nu },\left[ \stackrel{.}{x}^{\rho },x^{\mu }\right]
\right] -\left[ \stackrel{.}{x}^{\mu },\left[ \stackrel{.}{x}^{\rho },x^{\nu
}\right] \right] \right)  \label{gamma} \\
&&  \nonumber \\
&=&\frac{1}{2}\text{ }(\partial ^{\rho }g^{\nu \mu }-\partial ^{\nu }g^{\rho
\mu }-\partial ^{\mu }g^{\rho \nu })
\end{eqnarray}
Comparing with the usual law of motion of a particle in an electromagnetic
field, as in the situation of a flat space, we can put:

\begin{equation}
F^{\mu \nu }(x)=\frac{m}{q}\text{ }n^{\mu \nu }(x)
\end{equation}
and get the equation of motion of a particle in a curved space:

\begin{equation}
m\text{ }\frac{d\stackrel{.}{x}^{\mu }}{d\tau }=-m\text{ }\Gamma _{\nu \rho
}^{\mu }\stackrel{.}{x}^{\nu }\stackrel{.}{x}^{\rho }-q\text{ }F^{\nu \mu }%
\stackrel{.}{x}_{\upsilon }
\end{equation}
so that:

\begin{equation}
\left[ H,\stackrel{.}{x}^{\mu }\right] =-\text{ }\Gamma _{\nu \rho }^{\mu }%
\stackrel{.}{x}^{\nu }\stackrel{.}{x}^{\rho }-\frac{q}{m}\text{ }F^{\nu \mu }%
\stackrel{.}{x}_{\upsilon }
\end{equation}
Note the difference between the two tensor $N^{\mu \nu }$ and $F_{\mu \nu }$
whose definitions are:

\begin{equation}
\left\{ 
\begin{array}{c}
\left[ \stackrel{.}{x}^{\mu },\stackrel{.}{x}^{\nu }\right] =-\frac{N^{\mu
\nu }}{m}=-g^{\mu \rho }g^{\nu \sigma }\frac{N_{\rho \sigma }}{m} \\ 
\left[ \stackrel{.}{x}_{\mu },\stackrel{.}{x}_{\nu }\right] =-\frac{F_{\mu
\nu }}{m}=-g_{\mu \rho }g_{\nu \sigma }\frac{F^{\rho \sigma }}{m}
\end{array}
\right.
\end{equation}
and more generally:

\begin{equation}
\left\{ 
\begin{array}{c}
\left[ \stackrel{.}{x}^{\mu },f(\stackrel{.}{x},\tau )\right] =\frac{N^{\mu
\nu }}{m}\frac{\partial f(\stackrel{.}{x},\tau )}{\partial \stackrel{.}{x}%
^{\nu }} \\ 
\left[ \stackrel{.}{x}_{\mu },f(\stackrel{.}{x},\tau )\right] =\frac{F_{\mu
\nu }}{m}\frac{\partial f(\stackrel{.}{x},\tau )}{\partial \stackrel{.}{x}%
_{\nu }}
\end{array}
\right.
\end{equation}

\smallskip

As in the case of flat Minkowski space, it is not difficult to recover the
two groups of Maxwell equations with or without monopoles. In this last case
we must take the following definition for the dual field:

\begin{equation}
^{\ast }F^{\mu \nu }=\frac{1}{2\sqrt{-g}}\varepsilon ^{\mu \nu \rho \sigma
}F_{\rho \sigma }.
\end{equation}

Now we will show that the covariant derivative and the
curvature tensor can be naturally introduced with our formalism.

\subsection{Covariant derivative}

As in the flat-space case, the equation of motion can be rewritten in the
two following manners:

\begin{equation}
m\text{ }\frac{d\stackrel{.}{x}^{\mu }}{d\tau }=-m\text{ }\Gamma _{\nu \rho
}^{\mu }\stackrel{.}{x}^{\nu }\stackrel{.}{x}^{\rho }-q\text{ }F^{\nu \mu }%
\stackrel{.}{x}_{\upsilon }
\end{equation}
and:

\begin{equation}
m\text{ }\frac{d\stackrel{.}{x}^{\mu }}{d\tau }=m\text{ }\frac{\partial 
\stackrel{.}{x}^{\mu }}{\partial x^{\nu }}\stackrel{.}{x}^{\nu }
\end{equation}
we then put: 
\begin{equation}
\frac{\partial \stackrel{.}{x}^{\mu }}{\partial x^{\nu }}=-\Gamma _{\nu \rho
}^{\mu }\stackrel{.}{x}^{\rho }+\frac{q}{m}F^{\mu }{}_{\nu }=\left[
H^{\prime },\stackrel{.}{x}^{\mu }\right]  \label{vingt1}
\end{equation}
From equation (\ref{vingt1}), a covariant derivative can be defined by means
of the brackets. For an arbitrary vector we put:

\begin{equation}
m\text{ }\left[ \stackrel{.}{x}_{\nu },V^{\mu }(x)\right] =\frac{\partial
V^{\mu }(x)}{\partial x^{\nu }}
\end{equation}
We then define as the usual covariant derivative:

\begin{equation}
\left[ D_{\nu },V^{\mu }\right] =\frac{\partial V^{\mu }}{\partial x^{\nu }}%
+\Gamma _{\nu \rho }^{\mu }V^{\rho }
\end{equation}
and for an arbitrary mixed tensor:

\begin{equation}
\left[ D_{\nu },T^{\mu }{}_{\sigma }\right] =\frac{\partial T^{\mu
}{}_{\sigma }}{\partial x^{\nu }}+\Gamma _{\nu \rho }^{\mu }T^{\rho
}{}_{\sigma }-\Gamma _{\nu \sigma }^{\rho }T^{\mu }{}_{\rho }
\end{equation}
For the particular case of the velocity we get:

\begin{equation}
\left[ D_{\nu },\stackrel{.}{x}^{\mu }\right] =\frac{\partial \stackrel{.}{x}%
^{\mu }}{\partial x^{\nu }}+\Gamma _{\nu \rho }^{\mu }\stackrel{.}{x}^{\rho
}=\frac{q}{m}F^{\mu }{}_{\nu }
\end{equation}
and in addition we recover the standard result:

\[
\left[ D_{\nu },g^{\mu \nu }\right] =0 
\]

\subsection{Curvature tensor}

From this definition of the covariant derivative we can naturally express a
curvature tensor by means of the brackets. Let's compute the following
expressions:

\begin{eqnarray}
\left[ D^{\mu },\left[ D^{\nu },V^{\rho }\right] \right] &=&\left[ \stackrel{%
.}{x}^{\mu },\partial ^{\nu }V^{\rho }+\Gamma _{\sigma }^{\nu \rho
}V^{\sigma }\right] +\Gamma _{\alpha }^{\mu \nu }(\partial ^{\alpha }V^{\rho
}+\Gamma _{\sigma }^{\alpha \rho }V^{\sigma })+\Gamma _{\alpha }^{\mu \rho
}(\partial ^{\nu }V^{\alpha }+\Gamma _{\sigma }^{\alpha \nu }V^{\sigma }) 
\nonumber \\
&=&\partial ^{\mu }\partial ^{\nu }V^{\rho }+\partial ^{\mu }(\Gamma
_{\sigma }^{\nu \rho })V^{\sigma }+\Gamma _{\sigma }^{\nu \rho }(\partial
^{\mu }V^{\sigma })+\Gamma _{\alpha }^{\mu \nu }(\partial ^{\alpha }V^{\rho
}+\Gamma _{\sigma }^{\alpha \rho }V^{\sigma })  \nonumber \\
&&+\Gamma _{\alpha }^{\mu \rho }(\partial ^{\nu }V^{\alpha }+\Gamma _{\sigma
}^{\alpha \nu }V^{\sigma })
\end{eqnarray}
and therefore:

\begin{eqnarray}
\left[ D^{\mu },\left[ D^{\nu },V^{\rho }\right] \right] -\left[ D^{\nu
},\left[ D^{\mu },V^{\rho }\right] \right] &=&\partial ^{\mu }(\Gamma
_{\sigma }^{\nu \rho })V^{\sigma }-\partial ^{\nu }(\Gamma _{\sigma }^{\mu
\rho })V^{\sigma }+\Gamma _{\alpha }^{\mu \rho }\Gamma _{\sigma }^{\alpha
\nu }V^{\sigma }-\Gamma _{\alpha }^{\nu \rho }\Gamma _{\sigma }^{\alpha \mu
}V^{\sigma }  \nonumber \\
&&+\Gamma _{\alpha }^{\nu \mu }(\partial ^{\alpha }V^{\rho }+\Gamma _{\sigma
}^{\alpha \rho }V^{\sigma })-\Gamma _{\alpha }^{\mu \nu }(\partial ^{\alpha
}V^{\rho }+\Gamma _{\sigma }^{\alpha \rho }V^{\sigma })  \nonumber \\
&=&R^{\mu \nu \rho }{}_{\sigma }V^{\sigma }+\Omega ^{\mu \nu }{}_{\alpha
}D^{\alpha }V^{\rho }
\end{eqnarray}
where we have introduced the torsion tensor $\Omega ^{\mu \nu }{}_{\alpha }$ 
$=\Gamma _{\alpha }^{\nu \mu }-\Gamma _{\alpha }^{\mu \nu }=0,$ and the
curvature tensor $R^{\mu \nu \rho }{}_{\sigma }$ . Due to symmetric property
of the Christoffel symbols, the curvature tensor is reduced to:

\begin{equation}
R^{\mu \nu \rho }{}_{\sigma }V^{\sigma }=\partial ^{\mu }(\Gamma _{\sigma
}^{\nu \rho })V^{\sigma }-\partial ^{\nu }(\Gamma _{\sigma }^{\mu \rho
})V^{\sigma }+\Gamma _{\alpha }^{\mu \rho }\Gamma _{\sigma }^{\alpha \nu
}V^{\sigma }-\Gamma _{\alpha }^{\nu \rho }\Gamma _{\sigma }^{\alpha \mu
}V^{\sigma }
\end{equation}
The Jacobi identity gives: 
\begin{equation}
\left[ D^{\mu },\left[ D^{\nu },V^{\rho }\right] \right] +\left[ D^{\nu
},\left[ V^{\rho },D^{\mu }\right] \right] +\left[ V^{\rho },\left[ D^{\mu
},D^{\nu }\right] \right] =0
\end{equation}
that is: 
\begin{equation}
\left[ D^{\mu },\left[ D^{\nu },V^{\rho }\right] \right] -\left[ D^{\nu
},\left[ D^{\mu },V^{\rho }\right] \right] =\left[ \left[ D^{\mu },D^{\nu
}\right] ,V^{\rho }\right] =0
\end{equation}
and finally: 
\begin{equation}
\left[ \left[ D^{\mu },D^{\nu }\right] ,V^{\rho }\right] =R^{\mu \nu \rho
}{}_{\sigma }V^{\sigma }
\end{equation}

\smallskip

\textit{remark 2.} We can also define the Ricci and the electromagnetic
energy-impulsion tensors, but we were unable to deduce the Einstein equation
from this formalism. Naturally, we can write this equation with our brackets
as a constraint equation.

\textit{remark 3.} We can generalize the covariant derivative in including the 
skew-symmetric tensor $F^{\mu }{}_{\nu }$ in the definition. For this
we take into account the gauge curvature for the determination of the new
covariant derivative.
For a vectorial function of the velocity we write:
\begin{equation}
\left[ \Delta _{\nu },f^{\mu }(\stackrel{.}{x})\right] =\frac{\partial
f^{\mu }(\stackrel{.}{x})}{\partial x^{\nu }}+\Gamma _{\nu \rho }^{\mu
}f^{\rho }(\stackrel{.}{x})-\frac{q}{m}F{}_{\rho \nu }\frac{\partial f^{\mu
}(\stackrel{.}{x})}{\partial \stackrel{.}{x}_{\rho }}
\end{equation}

and then for the velocity:
\begin{equation}
\left[ \Delta _{\nu },\stackrel{.}{x}^{\mu }\right] =\frac{\partial 
\stackrel{.}{x}^{\mu }}{\partial x^{\nu }}+\Gamma _{\nu \rho }^{\mu }%
\stackrel{.}{x}^{\rho }-\frac{q}{m}F{}^{\mu }{}_{\nu }=0
\end{equation}

The covariant derivatives, are then simultaneously covariant under
both local internal and external gauges. If we want to keep a synthetic form
for the formulas using the curvature and torsion tensors, we must suppose for
an arbitrary vector the relation:
\begin{equation}
\left[ \Delta _{\nu },V^{\mu }\right] =\frac{\partial V^{\mu }}{\partial
x^{\nu }}+\Gamma _{\nu \rho }^{\mu }V^{\rho }-A_{\nu }V^{\mu }
\end{equation}
where the vector $A_{\nu }$ is defined by the following
equation: 
\begin{equation}
F^{\mu \nu }=m\left( \left[ \stackrel{.}{x}^{\mu },A^{\nu }\right] -\left[ 
\stackrel{.}{x}^{\nu },A\right] ^{\mu }\right)
\end{equation}
therefore we have: 
\begin{equation}
\left[ \Delta ^{\mu },\left[ \Delta ^{\nu },V^{\rho }\right] \right] -\left[
\Delta ^{\nu },\left[ \Delta ^{\mu },V^{\rho }\right] \right] =\left[ \left[
\Delta ^{\mu },\Delta ^{\nu }\right] ,V^{\rho }\right] =R^{\mu \nu \rho
}{}_{\sigma }V^{\sigma }+\Omega ^{\mu \nu }{}_{\alpha }\Delta ^{\alpha
}V^{\rho }+F^{\mu \nu }V^{\rho }
\end{equation}
We define a new ''generalized'' curvature tensor which matches the
local electromagnetism internal symmetry with the local external symmetry:
\begin{equation}
\stackrel{\_}{R}^{\mu \nu \rho }{}_{\sigma }V^{\sigma }=R^{\mu \nu \rho
}{}_{\sigma }V^{\sigma }+F^{\mu \nu }V^{\rho }
\end{equation}
then:
\begin{equation}
\left[ \left[ \Delta ^{\mu },\Delta ^{\nu }\right] ,V^{\rho }\right] =%
\stackrel{\_}{R}^{\mu \nu \rho }{}_{\sigma }V^{\sigma }
\end{equation}

\section{Conclusion}

The goal of this work was to study the dynamic associated with the 
Lorentz-covariant Hamiltonian well known in analytic mechanic. For this, we
introduced a four dimensional bracket structure which gives an algebraic 
structure between the position and velocity and generalizes the Poisson
brackets. This leads us to introduce a new time parameter which is not the
proper time, but is the conjugate coordinate of this covariant Hamiltonian.
This formal construction allows to recover the two groups of Maxwell
equations in flat space. This approach is close to the one used by Feynman
in his own derivation of the first group of Maxwell equations.

The principal interest of this method, besides the phase space formalism,
is in the study of theories with gauges symmetries because it avoids
the introduction of the non-gauge invariant momentum.

Our formalism can be directly extrapolated to the curved space,
where the principal notions are introduced in a natural manner. A 
five-dimensional structure can also be studied by considering the $\tau $ 
parameter as a fifth coordinate. In such a case equations take a simpler form,
particularly the group of Maxwell equations, but the meaning of this new
coordinate is still difficult to interpret, and could be perhaps understood
in the context of Kaluza-Klein compactification.

Just after finishing this work we received a paper referring to
the covariant Hamiltonian in the context of Feynman's proof of the Maxwell
equations\cite{MONTESINOS} .

\underline{Acknowledgment:} We would like to thank Y.Grandati for helpful
discussions.

\end{document}